\documentclass[prl,twocolumn,aps,showpacs]{revtex4}
\makeatletter
\def\@pnumwidth{2em}
\makeatother
\usepackage{graphicx}
\usepackage{dcolumn}
\usepackage{bm}
\def\be{\begin{equation}}
\def\ee{\end{equation}}
\begin{document}

\title{Universality of the metal-insulator transition in three-dimensional disordered systems}

\author{J. Brndiar and P. Marko\v{s}}%
\affiliation{Institute of Physics, Slovak Academy of
Sciences,  845 11 Bratislava, Slovakia}

\begin{abstract}
The universality of the metal-insulator transition in three-dimensional disordered system is  
confirmed by numerical  analysis of the scaling properties of the 
electronic wave functions. We prove that  the critical exponent  $\nu$
and the multifractal dimensions $d_q$ are 
independent on the microscopic definition of the disorder and
universal along the critical line  which separates  the metallic and the insulating regime.
\end{abstract}


\maketitle

One of the main problem of the disorder induced metal-insulator transition 
(MIT) is the proof of its universality. In the pioneering work \cite{AALR},  it was 
conjectured  that if the sample size exceeds all the length parameters
of the model, then  the conductance, $g$, is the only parameter which governs 
MIT. This scaling hypothesis has been confirmed by various numerical analysis,
with the help of the  finite-size scaling \cite{MacKK,angus1983}. 

\begin{figure}[b!]
\includegraphics[clip,width=0.45\textwidth]{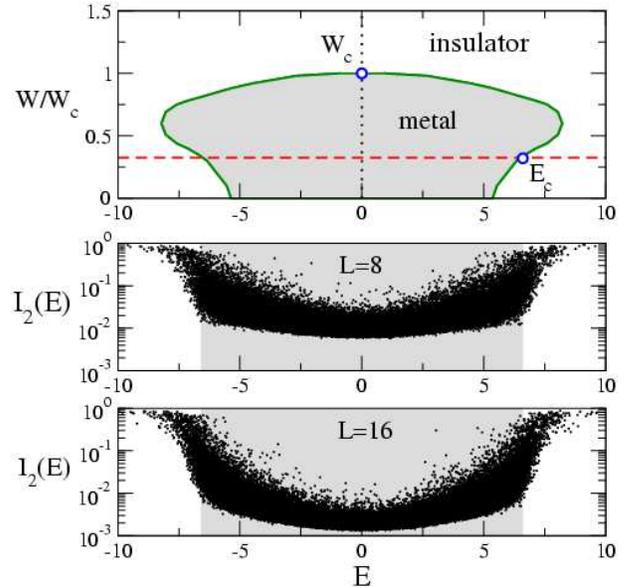}
\caption{(Color online) :w
Top panel shows schematic phase diagram for 3D Anderson model.
Solid line separates metallic (shaded area) and localized states.
Open circles shows the position of  critical points studied in this paper. 
Note that the mobility edge, $E_c$ lies outside the unperturbed energy band.
Two bottom panels 
present numerical data for $I_2$, given by Eq. (\ref{ipr})  for cubes  of 
the size $L=8$ (middle)
and $L=16$ (bottom) and with Gaussian disorder $W=2 = 0.325~W_c$. In the metallic region,
$I_2$ decreases when $L$ increases, while $I_2$ becomes $L$-independent in the tail
of the band, where electronic states are localized.
}
\label{schema}
\end{figure}

Generally accepted scenario of the Anderson localization is that  disorder
broadens the conductance band. Electron states in the tail of the band become
localized, separated from delocalized (metallic) states by the mobility edges, $E_c$.
System exhibits the MIT if the  Fermi energy, $E_F$,  crosses the mobility edge.
With increased disorder, $E_c$ moves towards the band center. There is a critical value
of the disorder, $W_c$, for which $E_c$ reaches band center, $E_c=0$. For disorder
$W>W_c$, all electronic states inside the band become localized. 
Phase diagram  
in the energy-disorder plane was calculated in \cite{bulka1985} and is schematically shown in the upper panel
of Fig. \ref{schema}.

At the band center, $E=0$, the universality of the MIT 
was confirmed by detailed numerical analysis
of the disorder and system size dependence of
Lyapunov exponents in quasi-one dimensional systems \cite{angus1983,SO99}, 
mean conductance \cite{SMO1},
conductance distribution \cite{SMO}, and  level statistics \cite{shklovskii1993,isa1995}.
These studies determined  the value of the  critical exponent $\nu$, which determines the divergence
of the correlation length, $\xi\sim|W-W_c|^{-\nu}$, as $\nu=1.57\pm 0.02$ \cite{SO99,SMO1}. 
The analysis of MIT along the critical line  (non-zero energy $E$) is more 
difficult because the critical region is narrower
and finite size effects are stronger \cite{kramer1990}. 
Critical exponent, $\nu$, was obtained only in
models with random hopping \cite{cain1999}, and very recently in \cite{croy2006}.

In this paper,  we present numerical proof of the universality of MIT.
By scaling analysis of   the electronic wave functions
in the vicinity of two critical points, shown in the upper panel of Fig. \ref{schema},
we prove that the critical exponent $\nu$ and fractal dimensions $d_q$ of the wave function
 (defined below) are universal along the critical line.

Electron eigenenergies and wave functions are calculated for three-dimensional
 Anderson Hamiltonian, 
\be\label{ham}
{\cal{H}}=W\sum_r \epsilon_r c_r^\dag c_r+\sum_{[rr']}c_r^\dag c_{r'}.
\ee
Here,  $r$ counts the sites of the three-dimensional (3D) lattice of the size $L^3$, 
$\epsilon_r$ is the random energy
distributed either with the Box distribution, $P_B(\epsilon)=(2/W)\Theta(W/2-|\epsilon|)$
or with the Gaussian distribution, $P_G(\epsilon)=\sqrt{2\pi W^2}\exp -(\epsilon^2/2W^2)$.
Parameter $W$ measures the strength of the disorder.
For $E=0$, the critical disorder
$W_c\approx 16.5$ ($6.15$) for the Box (Gauss)
distribution of random energies, respectively. 

The quantities of interest are
 inverse participation ratios (IPR),  $I_q(E_n)$. By definition, \cite{Kramer}
\be\label{ipr}
I_q(E_n)=\sum_{r} |\Phi_n(r)|^{2q}.
\ee
Here, $E_n$ and $\Phi_n(r)$  is the $n$th eigenenergy  and eigenfunction of the Hamiltonian (\ref{ham}),
respectively. 

The size dependence of IPR indicates the character of the eigenstate.
If the $n$th eigenstate is conductive, 
the wave function is distributed throughout the sample and
$|\Phi_n(r)|\propto L^{-d/2}$. Inserting in Eq. (\ref{ipr}) we obtain that 
$I_q(E_n)\propto L^{d(1-q)}$.
For localized state, $\Phi_n(r)$ is non-zero only in a small region, 
where $|\Phi_n(r)|\sim 1$. Hence, 
 $I_q(E_n)\sim 1$, too. These size dependences are shown 
in two bottom panels of Fig. \ref{schema}. 

Different size dependence of $I_q(E)$  enables us to use IPR as a scaling variable 
 for the calculation of the critical parameters in the same way as Lyapunov exponents
\cite{angus1983}, the conductance \cite{SMO1}, or level statistics \cite{shklovskii1993}.  
However, in contrast to the  Lyapunov exponent or mean conductance, IPR
is not a size independent constant at the critical point, but decreases as
\cite{evers2000,mirlin200a,milde2002} 
\be\label{ipr-crit}
I_q(E=E_c)\sim  L^{-d_q},
\ee
where $d_q$ are fractal dimensions.
This makes the scaling analysis slightly more  difficult. On the other hand,
fractal dimensions, $d_q$, represent a new set of   parameters, which can  be used for the
verification of the universality of the MIT. We expect that 
$d_q$  are universal constants for all critical points along the critical line.

\begin{figure}
\includegraphics[clip,width=0.35\textwidth]{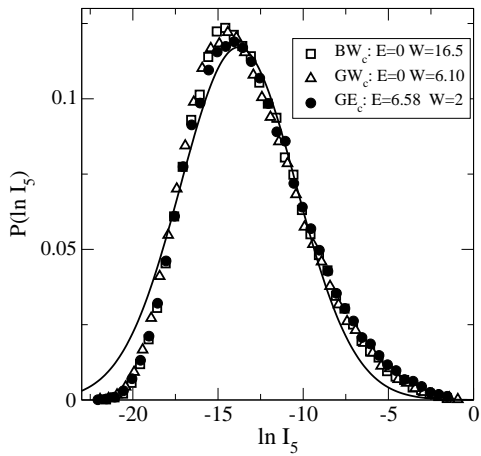}
\caption{Probability distribution $P(\ln I_5)$ for three critical points
and $L=40$. We used data  from  narrow energy interval, $E_c\pm\delta E$.
The size of the system is $L=40$.
Note that neither mean value nor the distribution is system-size independent.
Mean value, $\langle\ln I_5\rangle=-13.08$. The value of the  
variance, var $\ln I_5 = 11.6$, agrees with estimation of Ref.  \cite{milde2002}. 
Solid line is Gaussian distribution with the same mean value and variance.
}
\label{distr}
\end{figure}

The energy spectrum of the Hamiltonian depends on the system size, $L$, and on the microscopic
details of the disorder in a given sample. Therefore, we have to calculate an average values,
defined as follows. For each system 
size, we consider a statistical ensemble of $N_s$ samples which differ only in the realization
of random energies, $\epsilon_r$. For each sample, 
we calculate   all eigenenergies, $E_n$, lying  in a narrow
energy interval, $E-\delta,E+\delta$, and calculate corresponding $I_q(E_n)$. 
For the $i$th sample, the number of eigenstates, 
$n_i$, depends on the microscopic realization of the disorder. 
Also, since the values of $I_q$ might fluctuate in many orders in magnitude
in the critical region \cite{mirlin200a} (these fluctuations are shown   in Fig. \ref{distr}),
it is  more convenient to study logarithm of $I_q$. Thus, our scaling variable
is then defined as
\be\label{Y}
Y_q(E)=\displaystyle{\frac{1}{N_{\rm stat}}}\sum_i^{N_s}\sum_{|E-E_n|<\delta}\ln I_q(E_n),
\ee
where $N_{\rm stat}=\sum_i n_i$. In our calculations, 
$N_{\rm stat}\sim 10^5$ ($5\times 10^4$ for $L\ge 50$) and $\delta= 0.025$. With these 
parameters, we calculate $Y_q(E)$ with relative accuracy better than  $0.2\%$.
Numerical data were collected by LAPACK 
subroutines for $L\le 16$. For larger system size ($L\le 54$)
we use our own program based on the  Lanczos algorithm.

\begin{figure}[t]
\includegraphics[clip,width=0.45\textwidth]{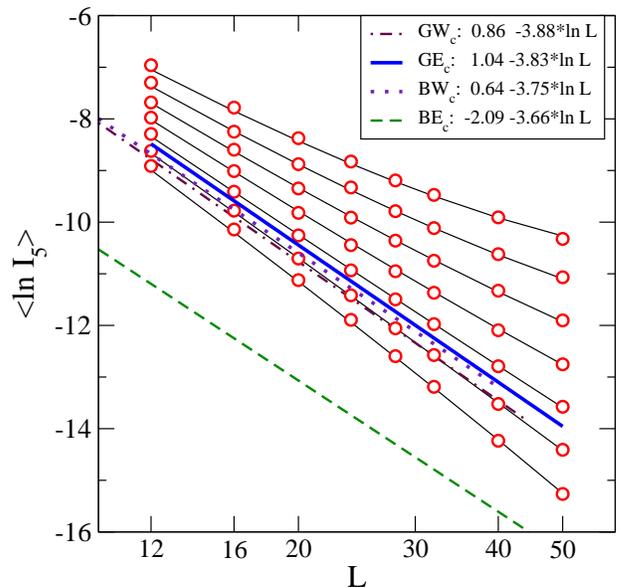}
\caption{(Color online) $Y_5(E,L)$ as a function of system size $L$ for various energies
$E= 6.50,~6.55,~\dots,6.80$ (from bottom to top). Thin solid lines are fits, Eq. (\ref{scaling-1}).
Thick solid line is $\langle\ln I_5\rangle=Y_5^c-4~d_5\ln L$ (Eq. \ref{xx})
 for the  critical energy $E_c\approx 6.58$. 
The same fit for other three critical points are also shown (described by legend).
Data confirm universal $L$-dependence of $Y_5$ for three critical points. The different behavior
for the critical point $BE_c$ is due to strong finite-size effect, discussed in the text.
}
\label{I5-final}
\end{figure}

\begin{table}[b]
\begin{tabular}{|l|l|ll|ll|ll|}
\hline
    & $q$  &    $E_c$              &   $W_c$ &    $\nu$            &  $d_q$ &  $L_{\rm min}$ & $L_{\rm max}$\\
\hline
\hline
$GW_c$    &  2   &  0       &   $6.14(3)$      &    $1.45(2)$  &  $1.28(2)$  &  16  &    44\\
$GW_c$    &  5   &  0       &   $6.07(4)$      &    $1.53(2)$  &  $0.97(8)$  &  16   &  44\\
$GE_c$    &  2   &  $6.59(1)$       &   2      &    $1.44(2)$  &  $1.28(4)$ &  20  &   50\\
$GE_c$    &  5   &  $6.58(3)$       &   2      &    $1.52(2)$  &  $0.96(5)$ &  20  &   50\\
\hline
$BW_c$        &  2   &  0       &   $16.70(10)$     &    $1.42(2)$  &  $1.23(7)$ &  16  &   40\\
$BW_c$         &  5   &  0       &   $16.53(10)$     &    $1.49(2)$  &  $0.93(9)$ &  16  &   40\\
$BE_c$         &  2   &  $7.44(2)$    &   10        &    $1.06(1)$  &  $1.11(8)$  &  32  &   54\\
$BE_c$         &  5   &  $7.43(3)$    &   10        &    $1.08(1)$  &  $0.87(32)  $  &  32  &   54\\
\hline
\end{tabular}
\caption{Critical exponent, $\nu$, and fractal dimensions, $d_q$, calculated by scaling 
analysis of the inverse participation ratio, $I_q$, for four critical points.
Calculated position
of critical points, $E_c,~W_c$, is given in the 3rd column.
$GW_c$ - Gaussian disorder, band center, $GE_c$ - Gaussian disorder, band tail,
$BW_c$ - Box disorder, band center, $BE_c$ - Box disorder, band tail. Data for system
of the size $L_{\rm min}\le L \le L_{\rm max}$ were used in the scaling analysis.
}
\label{table1}
\end{table}

We expect that $Y_q$ is a good scaling variable, so that it behaves in the vicinity of
the critical point as
\be\label{scaling-1}
Y_q(E,L)=Y_q^c-d_q\ln L +A(E-E_c)L^{1/\nu},
\ee
for the fixed disorder $W$, and 
\be\label{scaling-2}
Y_q(W,L)=Y_q^c-d_q\ln L +A(W-W_c)L^{1/\nu},
\ee
for the fixed  energy $E=0$.

\begin{figure}
\includegraphics[clip,width=0.35\textwidth]{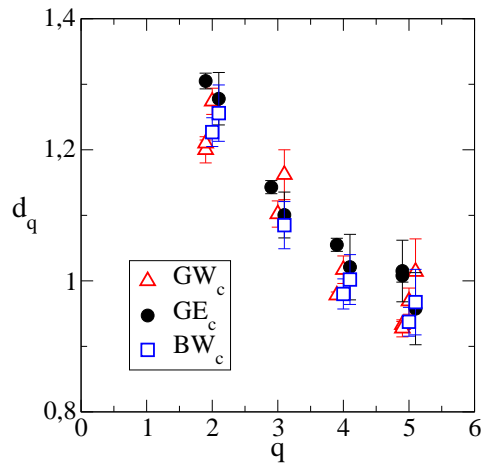}
\caption{(Color online) Fractal dimensions, $d_q$, calculated by scaling analysis of $Y_q$
for three critical points. 
 A small  horizontal shift of symbols for a given $q$  indicates 
minimal system size, $L_{\rm min}$, used in the scaling analysis:
 $L_{\rm min}= 8$ (left), 16 (middle) and 20 (right).  
}
\label{dq-final}
\end{figure}

\begin{figure}[t]
\includegraphics[clip,width=0.35\textwidth]{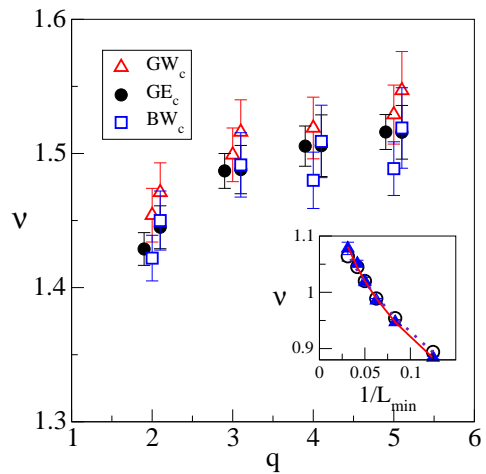}
\caption{(Color online) Critical exponent, $\nu$, 
calculated from the scaling behavior of $Y_q$ for three critical points. 
 A small  horizontal shift of symbols for a given $q$  indicates 
minimal system size, $L_{\rm min}$, used in the scaling analysis:
 $L_{\rm min}= 8$ (left), 16 (middle) and 20 (right).  
Inset shows the critical exponent $\nu$ calculated for critical point $BE_c$
from $Y_2$ (circles) and $Y_5$ (triangles). 
$\nu$ increases when data for smaller system size are omitted
($L_{\rm min}$ increases). This indicates that the deviation from universality, observed
in this critical point, is only finite size effect.
Solid (dashed) lines are fits $a_0+a_1/L^{a_2}$ with $a_0\approx 1.36$ (1.29) for $q=5$ ($q=2$),
respectively.
}
\label{nu-final}
\end{figure}

For a given $q$, we fit obtained numerical data for $Y_q$ to Eqs. (\ref{scaling-1})
or (\ref{scaling-2}). Obtained results are in Figs. \ref{I5-final}, \ref{dq-final} 
and \ref{nu-final}.
Typical data for $q=2$ and $q=5$ are given in Table \ref{table1}.

Figure \ref{I5-final} shows   the $L$-dependence of $Y_5$ for Gaussian disorder $W=2$
and for energies close to  the critical energy, $E_c\approx 6.58$. 
We see that at the critical point, $E=E_c$,   $Y_q(E_c)$
decreases logarithmically, 
\be\label{xx}
 Y_q(E_c,L)= Y_q^c -  d_q\ln L,
\ee
in agreement with Eq. (\ref{scaling-1}). 
Outside the critical point, the $L$ dependence of
$Y_q$ changes due to the presence of the term $\sim (E-E_c)L^{1/\nu}$. 
Similar analysis, performed  for other critical points confirms the universality
of the relationship (\ref{xx}).  This indicates that  parameters $Y_q^c$, $d_q$  and $A$ are
universal. 

There are two sources of inaccuracy of the scaling analysis: (i)
if the critical region is not sufficiently narrow, then the linear term $\sim E-E_c$
could not be sufficient to describe a correct  $E$-dependence of numerical data
and higher order terms of the expansion (\ref{scaling-1}) must be considered \cite{SO99}.
To test the accuracy of the   linear approximation, we 
 add also  cubic term, $\sim (E-E_c)^3$ in Eq. (\ref{scaling-1}).
 We found that such correction does not influence obtained critical parameters
 and might be neglected.
(ii) for small system  size, $L$, the variable $Y_q$  suffers from finite-size effects (FSE)
\cite{SO99}.
The role of FSE  can be  estimated by  the scaling analysis
for  data of restricted system size, $L\ge L_{\rm min}$. 
For three critical points, $GE_c$, $GW_c$  and $BW_c$ (the position of
critical points is given in Table \ref{table1}), we found that data for  $L>L_{\rm min}\sim 16-20$ 
are already free of FSE.   However, we were not able to obtain
reliable critical parameters for the 
 critical point $BE_c$.
As shown in Table \ref{table1} and Fig. \ref{I5-final}, our numerical data 
for this critical point are still
far from limiting values. For this critical point we can only demonstrate
the convergence of critical parameters to expected universal values when $L_{\rm min}$ increases
(inset of Fig. \ref{nu-final}). Larger samples are necessary to prove  this  convergence 
numerically.

Figure \ref{dq-final} summarizes our data for fractal dimensions, $d_q$. Data  confirm that the
spatial distribution of the wave function is universal, independent on the position of the
critical point along the critical line. Presented  data for $d_q$ are in agreement with
previously reported values \cite{milde2002}.

Figure  \ref{nu-final}
 presents obtained   data for the critical exponent, $\nu$.  
Wee see that $\nu$ converges to the generally accepted value, $\nu\sim 1.57$ \cite{SO99,SMO1},
when either $L_{\rm min}$ or $q$ increases. 
In order to show the effect of the  system size, we plot for each value of $q$ a few data
obtained with increasing minimal system size used in the scaling procedure.  
Inset of Fig. \ref{nu-final} presents our data for the critical point $BE_c$. Because of strong
finite-size effects, much larger systems are necessary for the estimation of reliable
values the critical exponent.

\begin{figure}
\includegraphics[clip,width=0.35\textwidth]{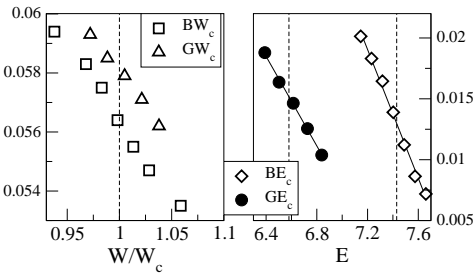}
\caption{The density of states, $\rho(E)$,
 arround the critical points. Dashed vertical lines indicates the position
of critical points.
Left: the disorder dependence of the density $\rho(E=0)$ at the band center.
$\rho(E=0)$ changes  only in 10\% 
when  disorder varies wihin the intervals ($15.5\le W\le 17.5$ for $BW_c$
and $6.5\le W\le 6.8$ for $GW_c$) used our scaling analysis. 
Left: The density of states, $\rho(E)$  around the critical points
$GE_c$ and $WE_c$.  
Solid lines are linear fits with slopes 0.018 ($GE_c$) and $0.026$ ($BE_c$).
}
\label{rho}
\end{figure}

To understand the origin of the FSE, we calculated the density of states, $\rho(E)$,
in the critical region of all four critical points.
As shown in left Fig. \ref{rho},  $\rho(E=0)$
changes only in a few \% when disorder varies around the critical value, $W_c$.
Contrary to the band center,  the density
$\rho(E)$ around the mobility edge $GE_c$ decreases significantly (almost by factor of two)
in the critical region (right Fig. \ref{rho}). 
In the case of critical point $BE_c$, this decrease is even more
significant.
Intuitively, one expects that the one-parameter scaling works  better when
electronic states inside  the critical region have the same, or at least comparable,
density.  
Since the interval of energies, $\Delta E=0.3$, used 
 in the case of $GE_c$ is already sufficient to get correct critical parameters,
 we expect   that 
the the  energy interval for $BE_c$ must be  $\Delta E\approx 0.2$. 
However, narrower interval of
energies requires larger system size. To have the term $(E-E_c)L^{1/\nu}$  of the same magnitude 
as in the $GE_c$-case, the size of the system must 
be $(3/2)\nu\approx 1.8$-times larger than in the $GE_c$ case.
Consequently, we expect that the scaling analysis of systems of the size  $L_{\rm max}\approx 90-100$
would lead to correct estimation of the critical exponent also for the critical point $BE_c$.

We expect that $I_q$  is less sensitive to finite size
effects  when $q$ increases.  Since the spatial distribution of electron is non-homogeneous in the 
critical regime, the main contribution to $I_q$ is given by sites with large values of  $|\Phi_n|$.
Those sites are well localized in space and so insensitive to the system size.
This is confirmed also  by  data shown in Fig. \ref{nu-final}.

The universality of the spatial distribution of the
critical wave function is confirmed also by Fig. \ref{distr} which shows
the probability distribution of $\ln I_5$ for three critical points, $GE_c$, $GW_c$ and $BW_c$.
The  width of the distribution,  $\sigma_q=\sqrt{{\rm var}\ln I_q}\sim 3.4$,
is close to the limiting value reported in Ref. \cite{evers2001}.
However, the distribution  $P(\ln I_q)$ is not system size invariant. 
The mean value of $\ln I_q$
decreases when system size increases, while the distribution always possesses a long tail for 
$\ln I_q\sim 1$ \cite{evers2001,cuevas,cuevas2002a}.

In conclusion, we investigated numerically the wave function of electron in the critical regime of 
the metal-insulator transition.  By the scaling analysis of the 
logarithm of the inverse participation ratio, $I_q$,
we calculated the critical exponent, $\nu$, and fractal dimensions,
$d_q$. We  found that these parameters depend neither on the microscopic details
of the disorder nor on the position of the critical point.
This result  confirms that the metal-insulator transition is universal along the critical line
which separates metallic and insulating regimes.

\medskip

This work was supported by grant APVV, project n. 51-003505 and VEGA, project n. 2/6069/26.


\begin{thebibliography}{99}


\bibitem{AALR} E.~Abrahams, P.~W.~Anderson, D.~C.~Licciardello, T.~V.~Ramakrishnan, Phys.~Rev.~Lett. {\bf 42}, 
673 (1979)

\bibitem{MacKK} A. MacKinnon,B. Kramer, Phys. Rev. Lett. {\bf 47} 1546 (1981)
\bibitem{angus1983} A. MacKinnon, B. Kramer, Z. Phys. B {\bf 53}, 1 (1983)


\bibitem{bulka1985} B. R. Bulka, B. Kramer, A. MacKinnon, Z. Phys. B {\bf 60}, 13 (1985)



\bibitem{SO99} K.~M.~Slevin, T.~Ohtsuki, Phys. Rev. Lett. {\bf 82}, 382 (1999)


\bibitem{SMO1} K.~Slevin, P.~Marko\v{s}, T.~Ohtsuki, Phys. Rev. Lett. {\bf 86}, 3594 (2001)

\bibitem{SMO} K.~Slevin, P.~Marko\v{s}, T.~Ohtsuki, Phys. Rev. B {\bf 67}, 155106 (2003)

\bibitem{shklovskii1993} B. I. Shklovskii, B. Shapiro, B. R. Sears, P. Lambrianides, B. H. Shore, Phys. Rev. B {\bf 47}, 11487 (1993)

\bibitem{isa1995} I. Kh. Zharekeshev, B. Kramer, Phys. Rev. B {\bf 51}, 17239 (1995)


\bibitem{kramer1990}  B.~Kramer, K.~Broderix, A.~MacKinnon, M.~Schreiber  Physica A {\bf 167}, 163 (1990)

\bibitem{cain1999} P. Cain, R. A. R\"omer, M. Schreiber, Ann. Phys. (Leipzig) {\bf 8}, SI-33 (1999)
\bibitem{croy2006} A. Croy, R. A. R\"omer, physica status solidi c {\bf 3}, 334 (2006)

\bibitem{Kramer} B.~Kramer, A.~MacKinnon, Rep. Progr. Phys. {\bf 56} 1469 (1993)



\bibitem{evers2000} F. Evers, A. D. Mirlin, Phys. Rev.Lett. {\bf 84}, 3690 (2000)
\bibitem{mirlin200a} A. D. Mirlin, F. Evers, Phys. Rev. B {\bf 62}, 7920 (2000)
\bibitem{milde2002} A. Mildenberger, F. Evers, A. D. Mirlin, Phys. Rev. B {\bf 66}, 033109 (2002)

\bibitem{evers2001} F. Evers, A. Mildenberger, A. D. Mirlin, Phys. Rev. B {\bf 64}, 241303(R) (2001)
\bibitem{cuevas} E. Cuevas, Phys. Rev. B {\bf 66}, 233103 (2002)

\bibitem{cuevas2002a} E. Cuevas, M. Ortuno, V. Gasparian, A. Perez-Garrido, Phys. Rev. Lett. {\bf 88}, 016401 (2002) 


\end{thebibliography}
\end{document}